\documentclass[%
 reprint,
preprintnumbers,
 amsmath,amssymb,
 aps,
prb,
]{revtex4-2}

\usepackage{orcidlink}

\usepackage{graphicx}
\usepackage{dcolumn}
\usepackage{bm}
\usepackage{hyperref}
\usepackage{amsmath,amssymb}
\hypersetup{colorlinks=true, linkcolor=blue, citecolor=blue, filecolor=blue,	urlcolor=blue}
\usepackage[normalem]{ulem}
\usepackage{makecell} 
\usepackage{placeins}
\begin{document}

\title{Unraveling the role of disorder in the electronic structure of high entropy alloys}

\author{Neeraj Bhatt\orcidlink{0009-0002-8012-139X}}
\affiliation{Department of Physics, Indian Institute of Science Education and Research Bhopal, Bhopal Bypass Road, Bhauri, Bhopal 462066, India}%

\author{Deepali Sharma\orcidlink{0009-0000-2309-4386}}
\affiliation{Department of Physics, Indian Institute of Science Education and Research Bhopal, Bhopal Bypass Road, Bhauri, Bhopal 462066, India}%

\author{Asif Ali\orcidlink{0000-0001-7471-8654}}
\affiliation{Department of Physics, Indian Institute of Science Education and Research Bhopal, Bhopal Bypass Road, Bhauri, Bhopal 462066, India}%

\author{Kapil Motla\orcidlink{0000-0002-0444-4632}}
\affiliation{Department of Physics, Indian Institute of Science Education and Research Bhopal, Bhopal Bypass Road, Bhauri, Bhopal 462066, India}%

\author{Sonika Jangid\orcidlink{0009-0003-7232-3757}}
\affiliation{Department of Physics, Indian Institute of Science Education and Research Bhopal, Bhopal Bypass Road, Bhauri, Bhopal 462066, India}%

\author{Ravi Prakash Singh\orcidlink{0000-0003-2548-231X}}
\affiliation{Department of Physics, Indian Institute of Science Education and Research Bhopal, Bhopal Bypass Road, Bhauri, Bhopal 462066, India}%

\author{Ravi Shankar Singh\orcidlink{0000-0003-3654-2023}}
\email{rssingh@iiserb.ac.in}
\affiliation{Department of Physics, Indian Institute of Science Education and Research Bhopal, Bhopal Bypass Road, Bhauri, Bhopal 462066, India}%

\date{\today}

\begin{abstract}

Disorder in high entropy alloys, arising from the random distribution of multiple elements, plays a crucial role in their novel properties desirable for various advanced engineering applications. We investigate the role of compositional and structural disorder on the electronic structure of osmium-based superconducting high entropy alloys, (Ru/Re)$_{0.35}$Os$_{0.35}$Mo$_{0.10}$W$_{0.10}$Zr$_{0.10}$, using photoemission spectroscopy and density functional theory (DFT). Elemental and cumulative core level shifts are found to be commensurate with elemental electronegativities and valence electron counts (VEC), respectively. Valence band spectra together with DFT results indicate that the crystal structure plays an important role in deciding the electronic structure of these high entropy alloys. Through temperature dependent high-resolution spectra, we unveil strongly suppressed spectral density of states (SDOS) in the close vicinity of Fermi level. Energy and temperature dependence of the SDOS in accordance with Altshuler-Aronov theory confirms localization of charge carriers in the presence of strong intrinsic disorder. Computed electron-phonon coupling strength and superconducting transition temperature aligning reasonably well with experiments further shed light on phonon-mediated pairing mechanism and role of disorder in these systems. Our results provide a way forward to the understanding of superconducting high entropy alloys through strategic control of disorder, VEC and crystal structure. 

\end{abstract}

\maketitle

High entropy alloys have garnered significant research interest in the area of material science over the past few decades due to extraordinary mechanical properties such as excellent strength, hardness, fracture toughness, and corrosion resistance, making them highly desirable for advanced engineering applications \cite{https://doi.org/10.1002/adem.200300567,doi:10.1126/science.1254581,Li2016,LI2019296}. The novel approach to metallic alloying by combining five or more elements in proportions ranging from 5 to 35 atomic \% results in high entropy alloys. The diverse composition of elements with substantially different atomic sizes results in a high degree of compositional disorder leading to high mixing entropy sets high entropy alloys apart from conventional alloys typically involving fewer elements with lower mixing entropy \cite{https://doi.org/10.1002/adem.200300567,ZHANG20141}. High entropy alloys have also emerged as unique class of materials due to their novel electronic, magnetic, and thermodynamic properties \cite{e15125338,MIRACLE2017448,MA201590,https://doi.org/10.1002/pssb.201800306,PhysRevB.96.014437}. Intrinsic disorder in high entropy alloys has profound implications on their electronic structure, affecting properties such as electron scattering, band broadening, and redistribution of spectral weight near the Fermi level ($E_F$) \cite{10.1063/1.5110519,NOH2000137,George2019,ma16041486,PhysRevMaterials.5.103604,Redka2024}. Thus, disorder plays a crucial role in determining transport, thermodynamic and mechanical properties and also the phase stability of high entropy alloys \cite{PhysRevLett.113.107001,Redka2024,PhysRevMaterials.5.103604,Xia2024}. High entropy alloys often exhibit novel properties distinct from their constituent elements, e.g., superconducting high entropy alloys exhibit significantly enhanced transition temperature ($T_c$) \cite{PhysRevMaterials.3.090301,PhysRevMaterials.3.060602,PhysRevB.105.144501,Zeng2024}. Also, the mechanism of superconductivity (SC) in high entropy alloys remains elusive due to limited research on the phononic and electronic band structures of these highly disordered multi-element compounds \cite{PhysRevLett.132.166002}.

Exploration of high entropy alloys have largely been centered around various 3$d$ or 4$d$ transition metals in diverse crystal structures such as face centered cubic (\textit{fcc}) \cite{10.1063/1.3587228}, body centered cubic (\textit{bcc}) \cite{10.1063/5.0200805}, hexagonal close packed (\textit{hcp}) \cite{PhysRevB.111.214504,PhysRevMaterials.3.060602}, orthorhombic (close to \textit{hcp}) \cite{LILENSTEN2014123} or in more complex noncentrosymmetric $\alpha$-\textit{Mn} \cite{PhysRevB.104.094515,Motla_2023} etc. Of these, the \textit{fcc} and \textit{bcc} high entropy alloys are by far the most extensively investigated having various 3$d$ and/or 4$d$ transition metals, primarily involving Fe/Co/Ni for magnetic-high entropy alloys \cite{ma14112877,Kitagawa2024} or Nb for SC-high entropy alloys \cite{Zeng2024,jangid2025,PhysRevB.104.094515,PhysRevLett.132.166002}.

Osmium-based high entropy alloys have recently been shown to exhibit SC with high $T_c$ ($\sim$ 5 times higher than average $T_c$ of its constituents), offering a platform to explore the role of crystallographic disorder on the electronic structure \cite{Motla_2023,Li_2025}. Ru$_{0.35}$Os$_{0.35}$Mo$_{0.10}$W$_{0.10}$Zr$_{0.10}$ (Ru-\textit{HEA}) having \textit{hcp} structure and Re$_{0.35}$Os$_{0.35}$Mo$_{0.10}$W$_{0.10}$Zr$_{0.10}$ (Re-\textit{HEA}) having $\alpha$-\textit{Mn} structure, exhibit $T_c$ of 2.90 K and 4.90 K, while the average $T_c$ is $\sim$ 0.55 K and $\sim$ 1.21 K, respectively \cite{Motla_2023}. Specific heat measurements suggest an isotropic $s$-wave SC gap in both systems while the upper critical fields exceed the Pauli limit by $\sim$ 30\%, hinting at possible unconventional SC \cite{Motla_2023}. 

In this Letter, we investigate the electronic structure of Ru-\textit{HEA} and Re-\textit{HEA} using high-resolution photoemission spectroscopy and density functional theory (DFT). Core level photoemission spectra reveal local charge redistribution governed by electronegativity differences of the constituent elements modulated by local atomic environments. Valence band spectra align well with DFT calculated electronic structure. Temperature dependent high-resolution spectra in the vicinity of  $E_F$ reveals strong suppression of electronic states suggesting localization of charge carriers in the presence of strong intrinsic disorder. The energy and temperature dependence of the spectral density of states (SDOS) follow Altshuler-Aronov theory for disordered systems \cite{ALTSHULER1979115,*RevModPhys.57.287,*PhysRevLett.98.246401}. Our computed electron-phonon coupling strength and $T_C$ together with the experimental results reveal the role of valence electron count (VEC) and compositional as well as intrinsic lattice disorder in determining
the superconducting properties of high entropy alloys.  

Photoemission spectroscopic measurements were performed on \textit{in-situ} scraped samples \cite{Motla_2023} using Scienta R4000 electron analyzer and monochromatic photon sources. Total instrumental resolutions were set to $\sim$ 280 meV for Al $K_{\alpha}$ (1486.6 eV) and $\sim$ 3 meV for He {\scriptsize I} (21.2 eV) radiations (energy). Electronic structure calculations were performed within DFT using QUANTUM ESPRESSO code \cite{Giannozzi_2009}, where optimized norm-conserving pseudopotentials \cite{PhysRevLett.43.1494,VANSETTEN201839} with the local density approximation \cite{PhysRevB.45.13244} were used. Atomic disorder in Ru-\textit{HEA} was modeled with a 60-atom special quasi-random structure supercell \cite{PhysRevLett.65.353}, while in Re-\textit{HEA} it was approximated by a random distribution of atoms in the 58-atom conventional unit cell. The phonon band structure, Eliashberg spectral function, electron-phonon coupling strength and $T_C$ were calculated using density functional perturbation theory (DFPT) \cite{RevModPhys.73.515}, within the virtual crystal approximation (VCA) \cite{PhysRevB.61.7877} for Ru-\textit{HEA} and Re-\textit{HEA} in \textit{hcp} structure. Further experimental and calculation methodologies are detailed in supplemental material SM \cite{supp}.

\begin{figure}[t]
    \centering
{\includegraphics[width=0.45\textwidth]{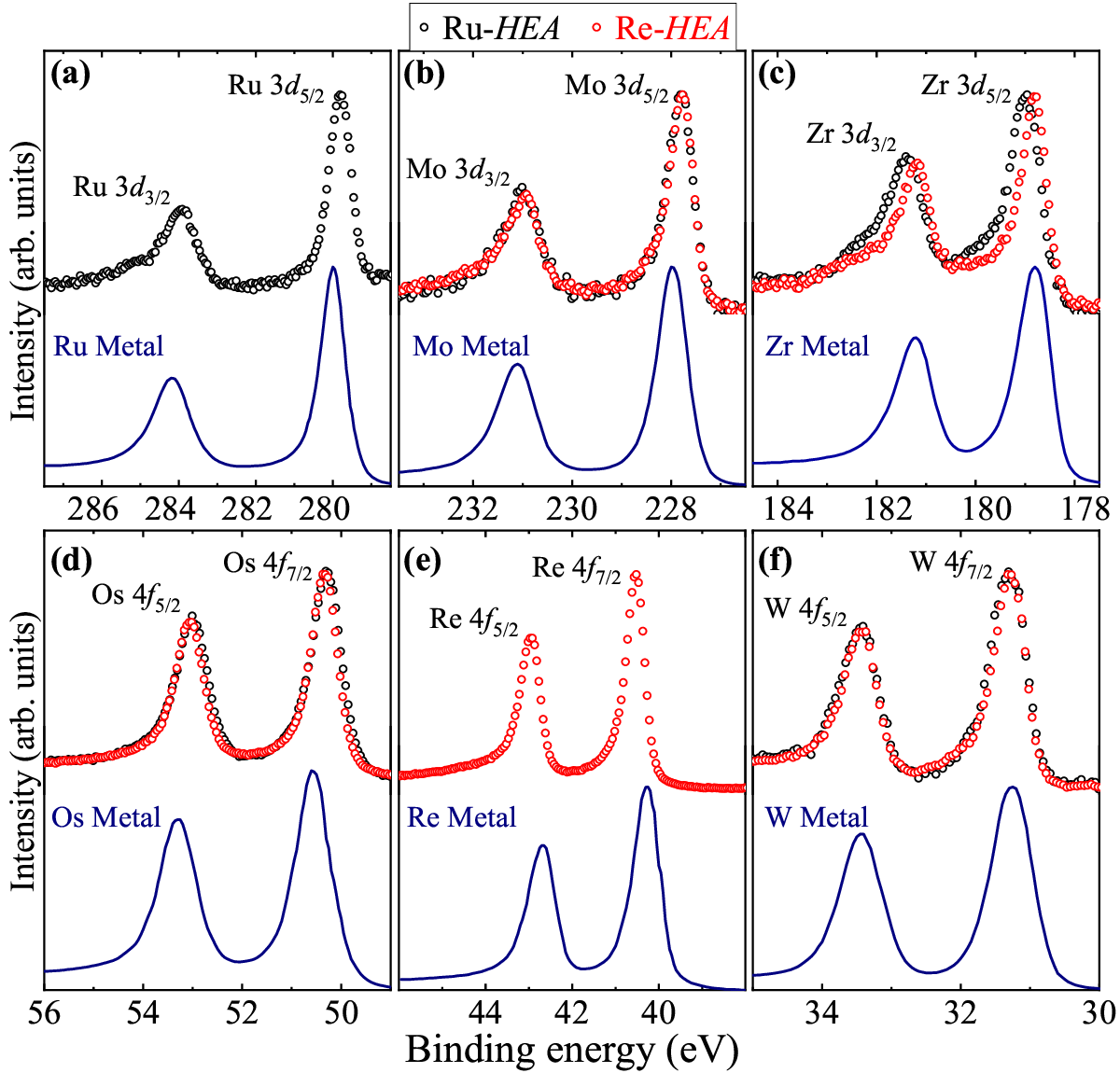}}    \caption{(a) Ru 3$d$, (b) Mo 3$d$, (c) Zr 3$d$, (d) Os 4$f$, (e) Re 4$f$, and (f) W 4$f$ core level photoemission spectra at 300 K for Ru-\textit{HEA} (black symbol) and Re-\textit{HEA} (red symbol). Spectra for elemental metals (blue line)  \cite{xpsdatabase} have also been shown.}  \label{Fig1}
\vspace{-4ex}
    \end{figure}

Ru-\textit{HEA} and Re-\textit{HEA}  crystallize in \textit{hcp} ($P6_3/mmc$, 194) and $\alpha$-\textit{Mn} ($I\Bar{4}3m$, 217) crystal structure respectively \cite{Motla_2023}. Core level spectral regions corresponding to Ru 3$d$, Mo 3$d$, Zr 3$d$, Os 4$f$, Re 4$f$, and W 4$f$ collected using Al $K_\alpha$ radiations at 300 K have been shown in Figs. \ref{Fig1}(a)$-$\ref{Fig1}(f) for Ru-\textit{HEA} (black) and Re-\textit{HEA} (red) [see Fig. S1 of SM for the survey scan \cite{supp}  ]. The spectra corresponding to the elemental metals \cite{xpsdatabase} have also been shown (blue lines) for comparison. As evident, all the core levels exhibit a spin-orbit split two-peak structure with Doniach-Sunjic type asymmetric line shape, indicative of highly metallic nature of the samples \cite{SDoniach_1970}. The binding energy (BE) positions of 3$d_{5/2}$ (for Ru, Mo, and Zr) and 4$f_{7/2}$ (for Os, Re, and W) peaks are summarized in Table \ref{Table1}, along with energy shift relative to the corresponding elemental core levels. Positive/negative  refers to the shift of the core levels toward higher/lower BE with respect to elemental metals.

\begin{table}[b]
\vspace{-4ex}
    \caption{Elemental core level positions and energy shifts (compared to metals) in Ru-\textit{HEA} and Re-\textit{HEA}.}
    \centering
    \renewcommand{\arraystretch}{1.5} 
    \resizebox{0.48\textwidth}{!}{%
    \begin{tabular}{|c|c|c|c|c|c|}
    \hline
        \textbf{Core level} & 
        \makecell{\textbf{Metal} \\ \textbf{position} \\ (eV)} & 
        \makecell{\textbf{Re-\textit{HEA}} \\ \textbf{position} \\ (eV)} & 
        \makecell{\textbf{BE} \\ \textbf{Shift} \\ (eV)} & 
        \makecell{\textbf{Ru-\textit{HEA}} \\ \textbf{position} \\ (eV)} & 
        \makecell{\textbf{BE} \\ \textbf{Shift} \\ (eV)} \\
    \hline
        Re 4$f_{7/2}$ & 40.26 & 40.58 & 0.32 & -- & -- \\
        Os 4$f_{7/2}$ & 50.52 & 50.32 & $-$0.20 & 50.27 & $-$0.25 \\
        W 4$f_{7/2}$  & 31.22 & 31.27 & 0.05 & 31.28 & 0.06 \\
        Mo 3$d_{5/2}$ & 227.91 & 227.79 & $-$0.12 & 227.79 & $-$0.12 \\
        Zr 3$d_{5/2}$ & 178.75 & 178.85 & 0.10 & 178.95 & 0.20 \\
        Ru 3$d_{5/2}$ & 279.95 & -- & -- & 279.75 & $-$0.20 \\
    \hline
    \end{tabular}%
    }
\label{Table1}
\end{table}

\begin{figure*}[t]
    \centering   {\includegraphics[width=0.9\textwidth]{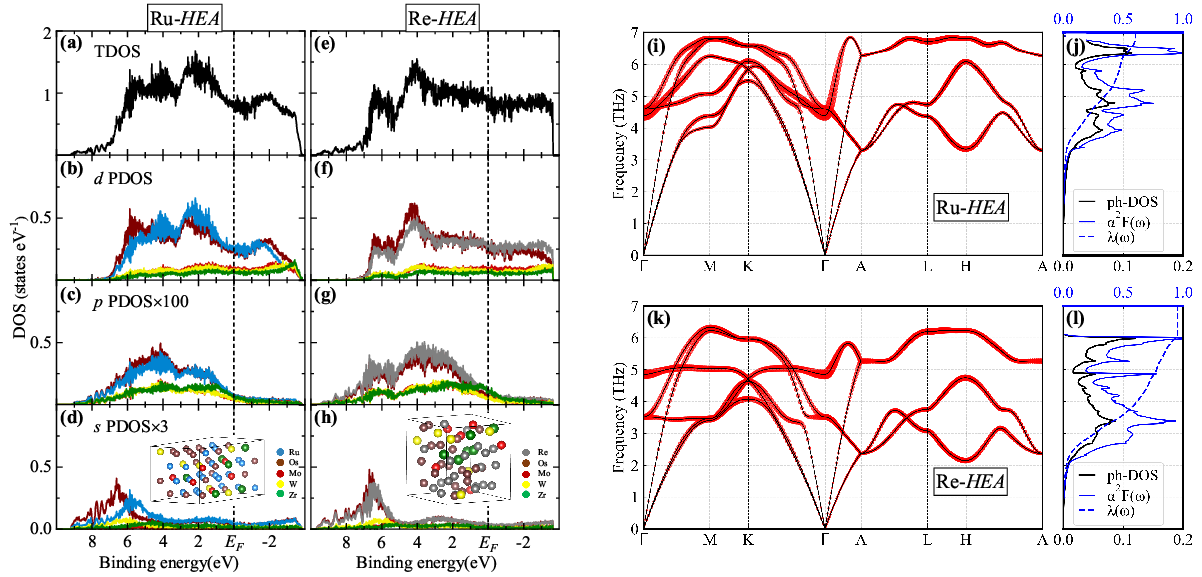}}
    \caption{ TDOS, $d$ PDOSs, $p$ PDOSs ($\times$100), and $s$ PDOSs ($\times$3) for Ru-\textit{HEA} are shown in (a), (b), (c), and (d), respectively, while the same for Re-\textit{HEA} are shown in (e), (f), (g), and (h) respectively. Insets of (d) and (h) show the 60-atom supercell for Ru-\textit{HEA} and 58-atom conventional cell for Re-\textit{HEA}, respectively. DFPT (within VCA) calculated phonon dispersion curves (lines) along with linewidth $\gamma_{\mathbf{q}\nu}$ (size of red symbols) in \textit{hcp} structure for (i) Ru-\textit{HEA} and (k) Re-\textit{HEA}. Ph-DOS, Eliashberg spectral function $\alpha^2\mathrm{F}(\omega)$, and $e\text{-}ph$ coupling strength $\lambda(\omega)$ for (j) Ru-\textit{HEA} and (l) Re-\textit{HEA}.} 
    \label{Fig2}
\vspace{-4ex}
\end{figure*}

The core level BE shift can be understood as interatomic electron transfer and relaxation effects arising from core-hole screening around the constituent atoms \cite{doi:10.1021/acsnano.0c09470,PhysRevLett.87.176403,STEINER19811885}. Thus, it is likely that, during alloy formation, some of the metal atoms act as electron acceptors (negative shift) or electron donors (positive shift) to the other metals \cite{doi:10.1021/acsnano.0c09470,https://doi.org/10.1111/jace.20021}. Os 4$f$ \& Mo 3$d$ and W 4$f$ \& Zr 3$d$ core levels shift towards lower and higher BE, respectively, suggesting that Os \& Mo act as electron acceptor with larger electronegativity while W \& Zr act as electron donor with relatively smaller electronegativity in both the samples. Ru 3$d$ in Ru-\textit{HEA} and Re 4$f$ in Re-\textit{HEA} exhibit lower (negative) and higher (positive) BE shifts corresponding to electron acceptor and  donor behaviour respectively. Our DFT calculated BE shifts for Ru 3$d$ core level in \textit{hcp} Ru-\textit{HEA} (w.r.t. Ru-metal) and for Re 4$f$ core level in $\alpha$-\textit{Mn} Re-\textit{HEA} (w.r.t. Re-metal) show the similar negative and positive shifts respectively [see Note III in SM \cite{supp}]. Interestingly, cumulative shifts of these core levels in Ru-\textit{HEA} and Re-\textit{HEA} are $-$0.31 eV and 0.15 eV respectively suggesting overall larger electron density in Ru-\textit{HEA} commensurate with larger VEC per atom of 7.2 as compared to 6.85 in Re-\textit{HEA}. Low temperature core level spectra collected at 30 K for all the elements in both the high entropy alloys remain almost identical to spectra collected at 300 K suggesting no change in the chemical or electronic properties [see Fig. S2 of SM \cite{supp}].

The results of electronic structure calculation within DFT for both high entropy alloys are shown in Fig. \ref{Fig2}. Total density of states (TDOS) along with atom projected partial DOS (PDOS) corresponding to $d$, $p$ and $s$ states for Ru-\textit{HEA} and Re-\textit{HEA} are shown in Figs. \ref{Fig2}(a)$-$\ref{Fig2}(d) and \ref{Fig2}(e)$-$\ref{Fig2}(h), respectively. Both the systems exhibit predominant contribution from $d$ states in 7 to $-$4 eV BE range as seen in Figs. \ref{Fig2}(b) and \ref{Fig2}(f), while $s$ states primarily appear at $\sim$ 6 eV BE as evident in Fig. \ref{Fig2}(d) and \ref{Fig2}(h) (scaled by $\times$3). Negligibly small contribution of $p$ states appear as broad features in the occupied energy range as shown in Figs. \ref{Fig2}(c) and \ref{Fig2}(g) (scaled by $\times$100). Insets of Figs. \ref{Fig2}(d) and \ref{Fig2}(h) show the fully relaxed unit cells for Ru-\textit{HEA} (60 atoms) and Re-\textit{HEA} (58 atoms), respectively, for which the DFT calculations were performed.
It is evident that the three discernible features at $\sim$ 6 eV, $\sim$ 4 eV along with a broad feature appearing below 3 eV binding energy in TDOS have significant contributions from Ru(Re) 4$d$(5$d$) states and Os 5$d$ states in Ru-\textit{HEA} (Re-\textit{HEA}). Both the systems exhibit a monotonously decreasing trend of TDOS with lowering BE in the vicinity of $E_F$ with significantly large intensity at $E_F$ representative of highly metallic nature of the samples. The apparent different in the spectral feature in the occupied as well as in unoccupied region of the two systems (differing by 0.35 VEC) can primarily be attributed to their structural differences. For comparison we compute the electronic structure of Ru-\textit{HEA} in $\alpha$-\textit{Mn} structure and Re-\textit{HEA} in \textit{hcp} structure which depicts the change in $E_F$ position corresponding to a rigid band shift due to the change in VEC in both the structures [See Fig. S3 in SM \cite{supp}]. The TDOS at $E_F$ are 0.83 and 0.88 states eV$^{-1}$ fu$^{-1}$ in close correspondence to the experimentally obtained values of 1.03 and 1.07 states eV$^{-1}$ fu$^{-1}$ for Ru-\textit{HEA} and Re-\textit{HEA} respectively.

Phonon band structure along with phonon linewidth, $\gamma_{\mathbf{q}\nu}$,  computed using DFPT \cite{RevModPhys.73.515} within VCA \cite{PhysRevB.61.7877} for Ru-\textit{HEA} and Re-\textit{HEA} in \textit{hcp} structure are shown in Figs. \ref{Fig2}(i) and \ref{Fig2}(k), while, phonon DOS (ph-DOS), Eliashberg spectral function, $\alpha^2\mathrm{F}(\omega)$, and electron-phonon ($e\text{-}ph$) coupling strength, $\lambda(\omega)$ are shown in Figs. \ref{Fig2}(j) and \ref{Fig2}(l) respectively. Cumulative $e\text{-}ph$ coupling strength $\lambda$ were used to evaluate the $T_C$ (details in SM \cite{supp}). For Ru-\textit{HEA}, obtained $\lambda$ (0.60) is in reasonable agreement with the experimental results of 0.52 while $T_C$ (3.77 K) is about 30\% higher than the experimental $T_C$ of 2.90 K, suggesting role of intrinsic disorder (absent in calculation) in suppressing the $T_C$, as also seen in other alloys \cite{PhysRevLett.132.166002}. Unlike explicit supercell approaches, VCA neglects local chemical environment and disorder effects and cannot account for  disorder-induced broadening or accurately reproduce high-frequency dispersions in phonon band structure \cite{Körmann2017}. Since, obtaining $\lambda$ and $T_C$ for Re-\textit{HEA} in the complex, low symmetry $\alpha$-\textit{Mn} structure with its 58 atoms unit cell requires forbiddingly expensive computation, we compute these quantities for Re-\textit{HEA} in \textit{hcp} structure resulting to $\lambda$ = 0.95 and $T_C$ = 9.50 K which are significantly larger than experimental values of 0.53 and 4.9 K, respectively, for $\alpha$-\textit{Mn} Re-\textit{HEA}. These results suggest that the crystal structure, intrinsic disorder along with VEC play a crucial role in determining the superconducting properties of these high entropy alloys. 

\begin{figure}[t]
    \centering   {\includegraphics[width=0.45\textwidth]{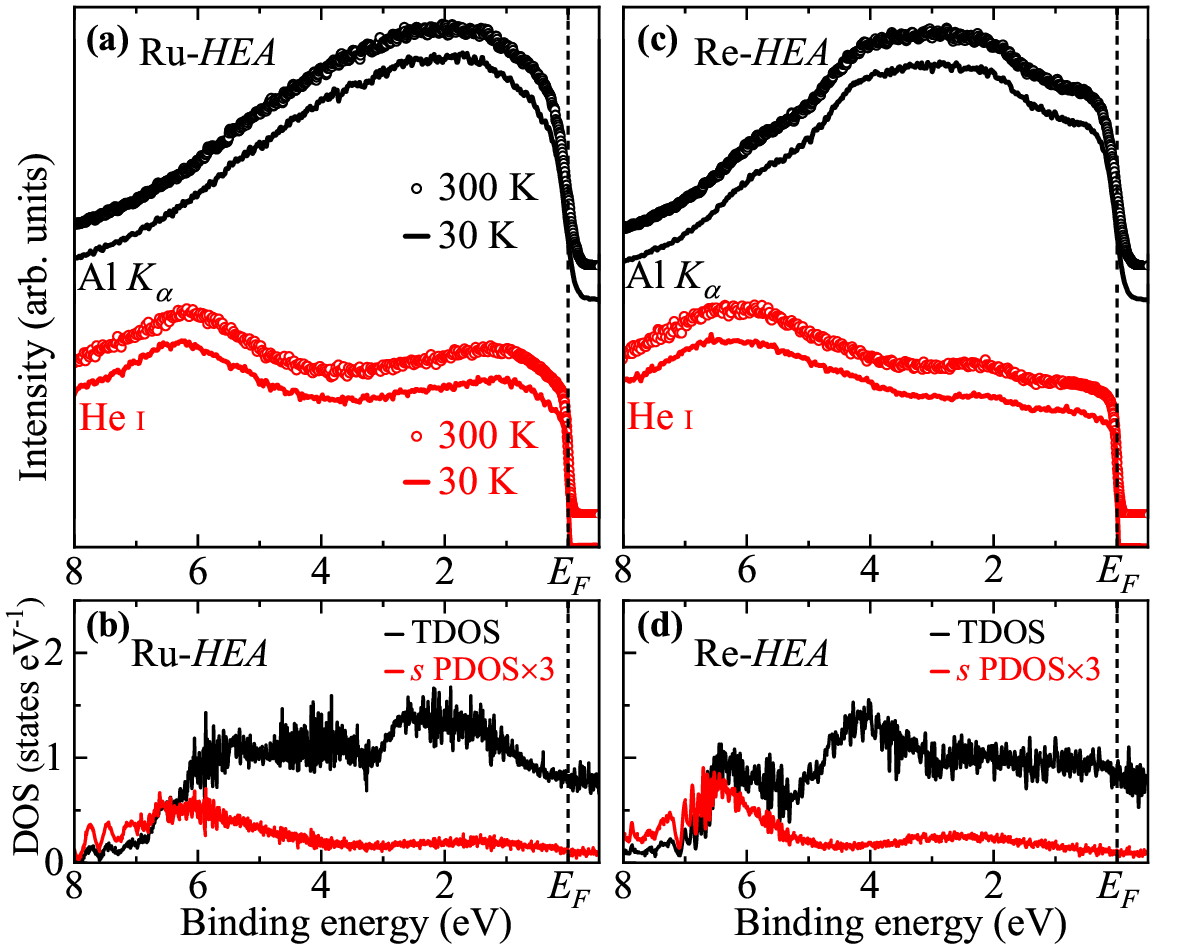}}
    \caption{Valence band spectra obtained using  Al $K_{\alpha}$ (black) and He {\scriptsize I} (red) at 300 K (symbols) and 30 K (lines) for (a) Ru-\textit{HEA} and (c) Re-\textit{HEA}. TDOS (black line) and $s$ PDOS (scaled $\times$3) for (b) Ru-\textit{HEA} and (d) Re-\textit{HEA}.} \label{Fig3}
\vspace{-4ex}
\end{figure}

To further understand the electronic structure, we show the valence band spectra collected using Al $K_{\alpha}$ and He {\scriptsize I} radiations at 300 K and 30 K for Ru-\textit{HEA} and Re-\textit{HEA} in Figs. \ref{Fig3}(a) and \ref{Fig3}(c) respectively. All the spectra are normalized by total integral intensity below 8 eV BE. For Ru-\textit{HEA}, Al $K_{\alpha}$ spectra at 300 K exhibits a broad feature around 2 eV BE with shoulder structures ranging upto $\sim$ 6 eV BE. A decreasing trend of the spectral intensity in the vicinity of $E_F$ along with a Fermi cutoff is clearly evident. Similarly, He {\scriptsize I} spectra also exhibit a broad structure $\sim$ 2 eV BE with a decreasing spectral intensity towards $E_F$ and a sharper Fermi cutoff due to higher energy resolution. Interestingly, He {\scriptsize I} spectra shows a broad structure centered $\sim$ 6 eV BE while it appears as a weak shoulder in Al $K_{\alpha}$ spectra.

The difference in the valence band spectra collected at two photon energies can primarily arise due to the different photoionization cross section of electronic orbitals involved in the formation of valence band \cite{YEH19851}. Thus, to understand the orbital contributions to the observed spectral change, we compare the valence band spectra with the TDOS (black) and cumulative PDOS of $s$-states (red), as shown in Fig. \ref{Fig3}(b) [see also Fig. \ref{Fig2}]. The prominent $s$-state contribution near 6 eV suggests its significance in the spectral feature. Although atomic photoionization cross section ratios [e.g., Ru (Os): $\sigma_{5s}/\sigma_{4d}$ ($\sigma_{6s}/\sigma_{5d}$) = 0.001 (0.002) at He {\scriptsize I} vs. 0.024 (0.057) at Al $K_{\alpha}$] indicate suppressed $s$-state contributions at lower photon energies \cite{YEH19851}, our results show otherwise. This discrepancy may arise from the Cooper minimum in $d$-states, which significantly suppresses their cross section at low photon energies, especially in high-Z atoms where relativistic effects and correlations in the form of interchannel coupling and/or configuration interaction in the final continuum states shift its energy position \cite{PhysRev.128.681,PhysRevB.28.3031,PhysRevLett.87.123004,PhysRevLett.46.1326,Cropper_2007}. The absence of any impurity-related signals and sharp core level spectra suggest the intrinsic nature of this feature, thus, we attribute the 6 eV feature to $s$-states. As shown in Fig. \ref{Fig3}(c), Al $K_{\alpha}$ spectra at 300 K of Re-\textit{HEA} exhibits a knee-line feature near $E_F$, followed by a broad flat region spanning from 2 eV to 4.5 eV BE and a shoulder structure around 6 eV BE. He {\scriptsize I} spectra at 300 K also reveals distinct features along with enhanced spectral weight at 6 eV BE corresponding of $s$ states similar to that observed in case of Ru-\textit{HEA}. Decreasing spectral intensity near $E_F$ along with Fermi cutoff is also evident in both the spectra for Re-\textit{HEA}. Similar to Ru-\textit{HEA}, valence band spectra of Re-\textit{HEA} are also in very well agreement with DFT results where the TDOS and cumulative PDOS of $s$ states are shown in Fig. \ref{Fig3}(d). 

\begin{figure}[t]
    \centering 
{\includegraphics[width=0.45\textwidth]{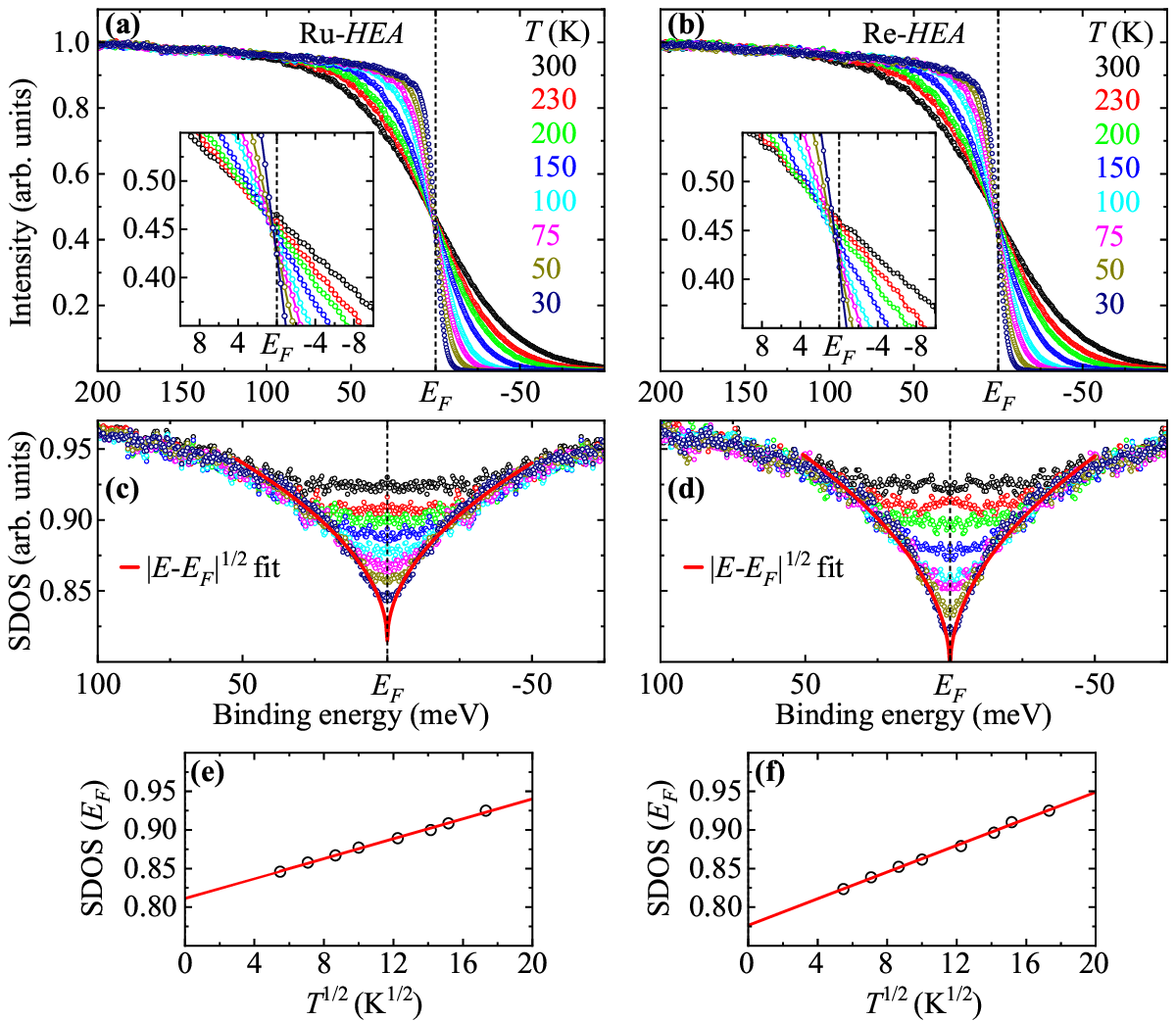}}
    \caption{ High-resolution He {\scriptsize I} spectra at different temperatures for (a) Ru-\textit{HEA} and (b) Re-\textit{HEA}. Inset shows the intensity in the vicinity of $E_F$. SDOS obtained using symmetrization of spectra for (c) Ru-\textit{HEA} and (d) Re-\textit{HEA} (red dashed line show the $\left|E-E_F\right|^{1/2}$ fit). $T^{1/2}$ behaviour of SDOS($E_F$) for (e) Ru-\textit{HEA} and (f) Re-\textit{HEA}. } \label{Fig4}
\vspace{-4ex}
\end{figure}

Similar to the elemental core levels, the valence band spectra collected using Al $K_{\alpha}$ as well as He {\scriptsize I} for both the high entropy alloys also remain almost identical in larger BE range while going from 300 K to 30 K (shown with lines and vertically shifted for clarity in Figs. \ref{Fig3}(a) and \ref{Fig3}(c)). It is to note that, apart from difference in the photoionization cross section the surface sensitivity also varies with photon energy where He {\scriptsize I} spectra is more surface sensitive than the Al $K_\alpha$ spectra. Nevertheless, the large intensity near $E_F$ region in the He {\scriptsize I} spectra (similar to Al $K_\alpha$ spectra) suggest that the states in the vicinity of $E_F$ are representative of bulk. Interesting to note that both the high entropy alloys exhibits a small decrease in spectral intensity at $E_F$ at low temperature. Since the electronic states in the close vicinity of $E_F$ play a dominant role in governing transport, thermodynamic, and various other physical properties of the system, we show the temperature dependent high-resolution spectra collected using He {\scriptsize I} radiations for Ru-\textit{HEA} and Re-\textit{HEA} in Figs. \ref{Fig4}(a) and \ref{Fig4}(b) respectively. All the spectra normalized at 200 meV BE exhibit a very similar lineshape with a decreasing trend upto $\sim$ 100 meV BE below which temperature dependent Fermi-Dirac like spectral evolution is clearly evident for both the systems. A closer look around $E_F$ reveals that the spectral intensity at $E_F$ monotonously decreases with decreasing temperature in case of Ru-\textit{HEA} as well as Re-\textit{HEA} as shown in the insets of Figs. \ref{Fig4}(a) and \ref{Fig4}(b) respectively.

For quantitative analysis of the temperature induced change, we obtain the SDOS by symmetrization of the photoemission intensity ($I(E)$) around $E_F$ \cite{PhysRevB.77.201102,PhysRevMaterials.7.064007,*Reddy_2019} and has been shown in Figs. \ref{Fig4}(c) and \ref{Fig4}(d). SDOS can also be obtained by dividing $I(E)$ with resolution broadened Fermi-Dirac function at respective temperatures \cite{PhysRevB.77.201102,PhysRevMaterials.7.064007,*Reddy_2019,PhysRevLett.95.016404} and expected to reveal similar spectral evolution provided that the SDOS does not abruptly change in the vicinity of $E_F$ (details in SM \cite{supp}). Almost identical SDOS, obtained using different methods, provides confidence in the analysis [see Fig. S4 in SM \cite{supp}]. The temperature dependent SDOS exhibit linearly decreasing trend and remain similar down to $\sim$ 50 meV BE below which there appears a cusp like feature around $E_F$ which evolves with decreasing temperature for both the systems. The evolution of dip like structure along with monotonic decrease of SDOS($E_F$) with decreasing temperature suggest the localization of electrons at $E_F$ in the presence of strong disorder \cite{PhysRevMaterials.7.064007,*Reddy_2019}. Interestingly, the SDOS(E$_F$) decreases from 0.92 to 0.85 for Ru-\textit{HEA} and from 0.92 to 0.82 for Re-\textit{HEA} while going from 300 K to 30 K. Larger suppression of SDOS($E_F$) in case of Re-\textit{HEA} ($\sim$ 11\%) than that in case of Ru-\textit{HEA} ($\sim$ 7.5\%) for the same compositional disorder suggests that the crystal structural also plays an important role where $\alpha$-\textit{Mn} Re-\textit{HEA} is intrinsically disordered than \textit{hcp} Ru-\textit{HEA}. The $\left|E-E_F\right|^{1/2}$ lineshape of the SDOS could be nicely captured for 50 meV $> \left|E-E_F\right| > k_BT$ for all the temperatures as shown by the red lines in Figs. \ref{Fig4}(c) and \ref{Fig4}(d). Also, the temperature dependent SDOS($E_F$) showcases $T^{1/2}$ behaviour for both the systems as depicted is Figs. \ref{Fig4}(e) and \ref{Fig4}(f) where the red lines show the linear fit to the data. $\left|E-E_F\right|^{1/2}$ behaviour of the SDOS and the square-root temperature dependence of SDOS($E_F$) are in agreement with the Altshuler-Aronov theory \cite{ALTSHULER1979115,*RevModPhys.57.287,*PhysRevLett.98.246401}, confirming disorder induced localization of electronic states and its significant role in determining the electronic structure of Ru-\textit{HEA} and Re-\textit{HEA}. 

In conclusion, our combined photoemission spectroscopy and DFT results reveal how  disorder, charge redistribution, and crystallographic complexity govern the electronic structure of high entropy alloys. Core level shifts highlight element-specific charge redistribution, modulated by local symmetry and bonding environments in different structures. High resolution spectra in the vicinity of $E_F$ reveals strong suppression of SDOS with decreasing temperature. Square-root energy and temperature dependence of SDOS and SDOS($E_F$), respectively, follows the Altshuler-Aronov theory for disordered systems indicating strong localization of electronic states in the presence of structural as well as compositional disorder in these high entropy alloys. Computed $e\text{-}ph$ coupling and $T_C$ aligning reasonably well with the experimental observations validate the role of phonon-mediated pairing in these systems and also indicate the role of disorder, VEC and structure in determining the superconducting propeties. Our work provides a way forward to the understanding of disordered superconductors through strategic control of disorder and crystal structure. 

\section{ACKNOWLEDGMENTS}
N.B. and D.S. acknowledge the CSIR India, for financial support through Awards No. 09/1020(0177)/2019-EMR-I and 09/1020(0198)/2020-EMR-I, respectively. R.P.S. acknowledges the SERB India, for Core Research Grant No. CRG/2023/000817. We gratefully acknowledge
the use of the HPC facility and CIF at IISER
Bhopal.

\end{document}


\title{Supplemental Material for ``Unraveling the role of disorder in the electronic structure of high entropy alloys"}

\author{Neeraj Bhatt\orcidlink{0009-0002-8012-139X}}
\affiliation{Department of Physics, Indian Institute of Science Education and Research Bhopal, Bhopal Bypass Road, Bhauri, Bhopal 462066, India}%

\author{Deepali Sharma\orcidlink{0009-0000-2309-4386}}
\affiliation{Department of Physics, Indian Institute of Science Education and Research Bhopal, Bhopal Bypass Road, Bhauri, Bhopal 462066, India}%

\author{Asif Ali\orcidlink{0000-0001-7471-8654}}
\affiliation{Department of Physics, Indian Institute of Science Education and Research Bhopal, Bhopal Bypass Road, Bhauri, Bhopal 462066, India}%

\author{Kapil Motla\orcidlink{0000-0002-0444-4632}}
\affiliation{Department of Physics, Indian Institute of Science Education and Research Bhopal, Bhopal Bypass Road, Bhauri, Bhopal 462066, India}%

\author{Sonika Jangid\orcidlink{0009-0003-7232-3757}}
\affiliation{Department of Physics, Indian Institute of Science Education and Research Bhopal, Bhopal Bypass Road, Bhauri, Bhopal 462066, India}%

\author{Ravi Prakash Singh\orcidlink{0000-0003-2548-231X}}
\affiliation{Department of Physics, Indian Institute of Science Education and Research Bhopal, Bhopal Bypass Road, Bhauri, Bhopal 462066, India}%

\author{Ravi Shankar Singh\orcidlink{0000-0003-3654-2023}}
\email{rssingh@iiserb.ac.in}
\affiliation{Department of Physics, Indian Institute of Science Education and Research Bhopal, Bhopal Bypass Road, Bhauri, Bhopal 462066, India}%

\date{\today}

 \maketitle

 \section{ Experimental Methodology}

Stoichiometric ratios of high-purity (99.99 \%) transition metals Re, Ru, Os, Mo, W, and Zr were arc melted using a standard single-arc melter in an argon atmosphere at 1.013 bar pressure to synthesize Ru$_{0.35}$Os$_{0.35}$Mo$_{0.10}$W$_{0.10}$Zr$_{0.10}$ (Ru-\textit{HEA}) and Re$_{0.35}$Os$_{0.35}$Mo$_{0.10}$W$_{0.10}$Zr$_{0.10}$ (Re-\textit{HEA}). A Ti getter was melted first to eliminate any remaining oxygen. The ingots were  remelted 5-6 times with intermediate flipping to ensure homogeneity and subsequently annealed at 1100 $^\circ$C for 7 days in a vacuum-sealed quartz tube. Detailed structural characterization and bulk measurements have been reported elsewhere \cite{Motla_2023}.
Temperature dependent photoemission spectroscopic measurements were performed on \textit{in situ} scraped samples using Scienta R4000 electron analyzer and monochromatic photon sources. Total instrumental resolutions were set to $\sim$ 280 meV for Al $K_{\alpha}$ (1486.6 eV) and $\sim$ 3 meV for He {\scriptsize I} (21.2 eV) radiations (energy). Clean polycrystalline silver was used to determine the Fermi energy ($E_F$) and the energy resolutions for different radiations at 30 K. The error in estimation of $E_F$ is about 0.2 meV. The reproducibility of data and cleanliness of the sample surface were insured after each trails of scraping by minimizing the intensity corresponding to C 1$s$ and O 1$s$ signals. 

 \section{ Survey scan and temperature dependent core level spectra}

\begin{figure*}[t]
\centering
{\includegraphics[width=1\textwidth]{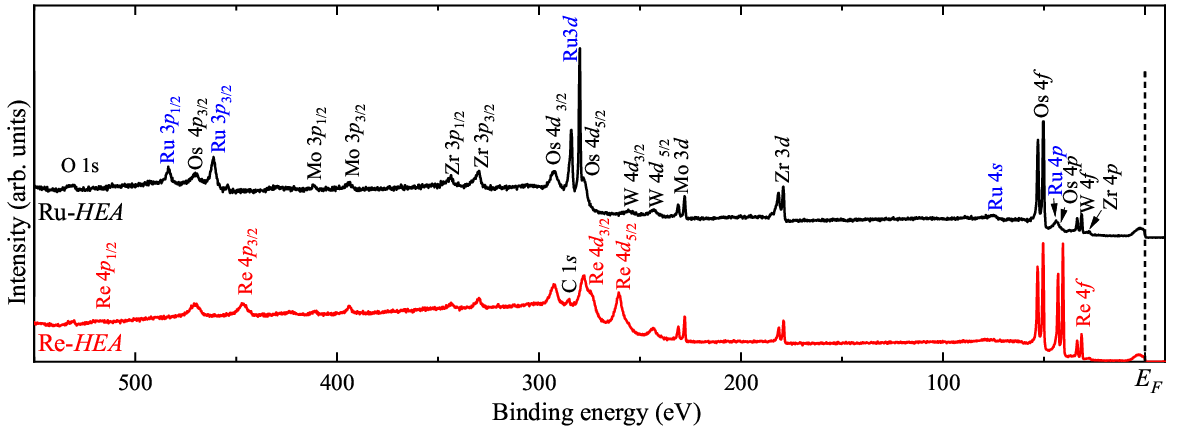}}
    \caption{Survey scan of Ru-\textit{HEA} (black) and Re-\textit{HEA} (red) are vertically stacked. Ru and Re core levels are marked in blue and red respectively, while others have been marked in balck.}\label{FigS1}
\vspace{8ex}
    \end{figure*}

Figure \ref{FigS1} show the survey scan collected using Al $K_{\alpha}$ radiation for Ru-\textit{HEA} (black) and Re-\textit{HEA} (red). Both the spectra are vertically stacked for clarity and all the elemental core levels are labeled in black color. Core levels corresponding to Ru and Re atoms have been specifically marked in blue and red colors, respectively. Negligibly small intensity corresponding to oxygen and carbon features in both the sample confirm clean sample surface. 

\begin{figure*}[t]
\centering
{\includegraphics[width=.98\textwidth]{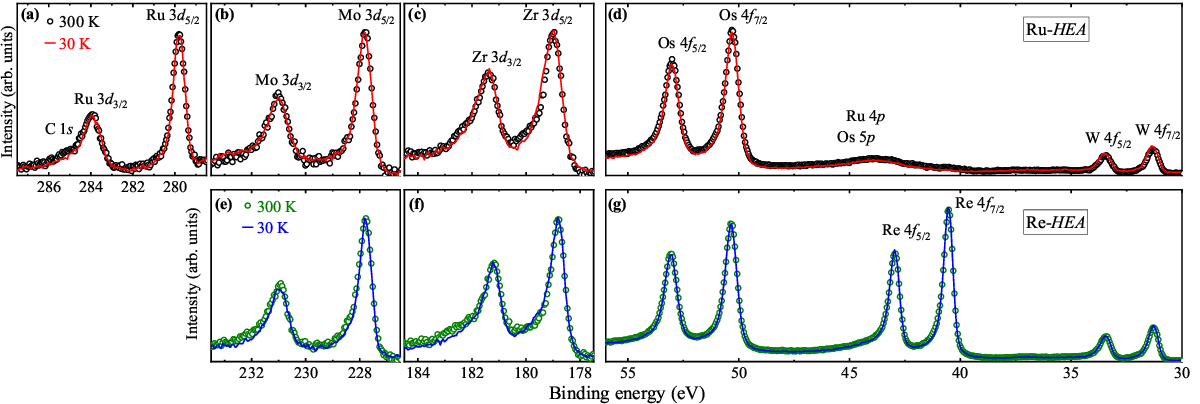}}
    \caption{Various elemental core level spectra (a)$-$(d) for Ru-\textit{HEA}, and (e)$-$(g) for Re-\textit{HEA} at 300 K (symbol) and 30 K (lines).} \label{FigS2}
\vspace{0ex}
    \end{figure*}

Core level spectra corresponding to the Ru 3$d$, Mo 3$d$, Zr 3$d$, and Os 4$f$ + W 4$f$ for Ru-\textit{HEA} at 300 K and 30 K are shown in Figs. \ref{FigS2}(a)$–$\ref{FigS2}(d), while the Mo 3$d$, Zr 3$d$, and Os 4$f$ + Re 4$f$ + W 4$f$ core level spectra for Re-\textit{HEA} at 300 K and 30 K are shown in Fig. \ref{FigS2}(e)$–$\ref{FigS2}(g). For both the samples all the core level spectra remain almost identical while going from 300 K to 30 K. The absence of temperature dependent changes in the core levels and in the valence band confirm that the disorder induced effects are only confined to electronic states in the close vicinity of the $E_F$, consistent with previous reports on disordered alloys and complex materials, where enhanced electronic localization  affects only the states close to $E_F$, while leaving the deeper electronic structure unchanged \cite{NOH2000137,PhysRevMaterials.7.064007,*Reddy_2019}.

\section{ Computational Methodology}

Electronic structure calculations were performed within density functional theory (DFT) using QUANTUM ESPRESSO (QE) \cite{Giannozzi_2009}. The optimized norm-conserving pseudopotentials \cite{PhysRevB.88.085117,VANSETTEN201839} with the local density approximation \cite{PhysRevB.45.13244} were used with wavefunctions and charge density cutoffs of 140 Ry and 1400 Ry, respectively. The valence configuration includes 4$s$, 4$p$, 4$d$ and 5$s$ states for Zr, Mo and Ru, and 5$s$, 5$p$, 5$d$ and 6$s$ states for W, Re, and Os, while the remaining lower lying states are treated as frozen core in corresponding pseudopotentials. To accommodate the random arrangement of atoms in the Ru-\textit{HEA} (\textit{hcp}) with appropriate stoichiometry, we use 5$\times$3$\times$2 supercell with special quasi-random structure (SQS) \cite{PhysRevLett.65.353} having 60 atoms (21 for Ru and Os and 6 for Mo, W, and Zr). The SQS supercell was constructed using the Monte Carlo Special Quasirandom Structure code as implemented in the Alloy Theoretic Automated Toolkit package \cite{VANDEWALLE201313}. In generating SQS, pair and triplet correlations were considered up to 5 \AA. The conventional unit cell contains 58 atoms for Re-\textit{HEA} (noncentrosymmetric $\alpha$-\textit{Mn}) and the smallest commensurate SQS supercell that exactly reproduces the target stoichiometry would consist of 10-unit cells (580 atoms). Thus to avoid prohibitively large computation we used randomly distributed atoms (20 for Re and Os and 6 for Mo, W, and Zr) in 58-atom conventional unit cell closely representing the actual compositional ratio. Various structures for Re-\textit{HEA} (with different atomic distribution) were generated and the electronic structure were found to be almost identical (not shown here). Full structural relaxations were performed with forces and energy convergence better than 10$^{-5}$ Ryd/Bohr and 10$^{-6}$ Ryd, respectively. 3$\times$5$\times$4 (Ru-\textit{HEA}) and 5$\times$5$\times$5 (Re-\textit{HEA}) \textit{k}-mesh were used for all the calculations. The tetrahedron method with Bl\"ochl corrections \cite{PhysRevB.49.16223} was employed for calculating electronic DOS. The phonon band structure, phonon lifetime broadening, Eliashberg spectral function, electron-phonon coupling strength and $T_C$ were calculated using density functional perturbation theory (DFPT) \cite{RevModPhys.73.515}, within the virtual crystal approximation (VCA) \cite{PhysRevB.61.7877} for Ru-\textit{HEA} and Re-\textit{HEA} in \textit{hcp} structure. The pseudopotential for a virtual atom representing the alloy composition Ru$_{0.35}$Os$_{0.35}$Mo$_{0.10}$W$_{0.10}$Zr$_{0.10}$ and Re$_{0.35}$Os$_{0.35}$Mo$_{0.10}$W$_{0.10}$Zr$_{0.10}$ were generated using the \texttt{Virtualv2.x} utility in QE. A 9$\times$9$\times$5 $q$-mesh and 18$\times$18$\times$10 $k$-mesh were employed for these calculations. The energy threshold of $10^{-14}$ Ryd was used for phonon calculations while broadening parameter of 0.02 Ryd was used to obtain the Eliashberg spectral function, electron-phonon coupling strength and superconducting transition temperature $T_C$. It is to note that the phonon calculations for the Re-\textit{HEA} in $\alpha$-\textit{Mn} structure are computationally prohibitive even with VCA, due to the large unit cell (29 atom for primitive cell).

The binding energies (BEs) of the Ru 3$d$ and Re 4$f$ core levels were computed using DFT within the full potential linearized augmented plane wave method, as implemented in WIEN2k \cite{10.1063/1.5143061}. For Ru-metal and Ru-\textit{HEA} in the \textit{hcp} structure, the obtained Ru 3$d$ core level BEs are 268.35 eV and 268.00 eV, respectively, suggesting a negative shift of $-0.35$ eV, while for Re-metal and Re-\textit{HEA} in the $\alpha$-\textit{Mn} structure, the Re 4$f$ core level BEs are 35.90 eV and 36.10 eV, respectively, exhibiting a positive shift of 0.20 eV. These core level shifts are in the same energy direction and of very similar amounts to those observed experimentally (see Table I in the main text). 

 \begin{figure*}[t]
    \centering
{\includegraphics[width=0.9
\textwidth]{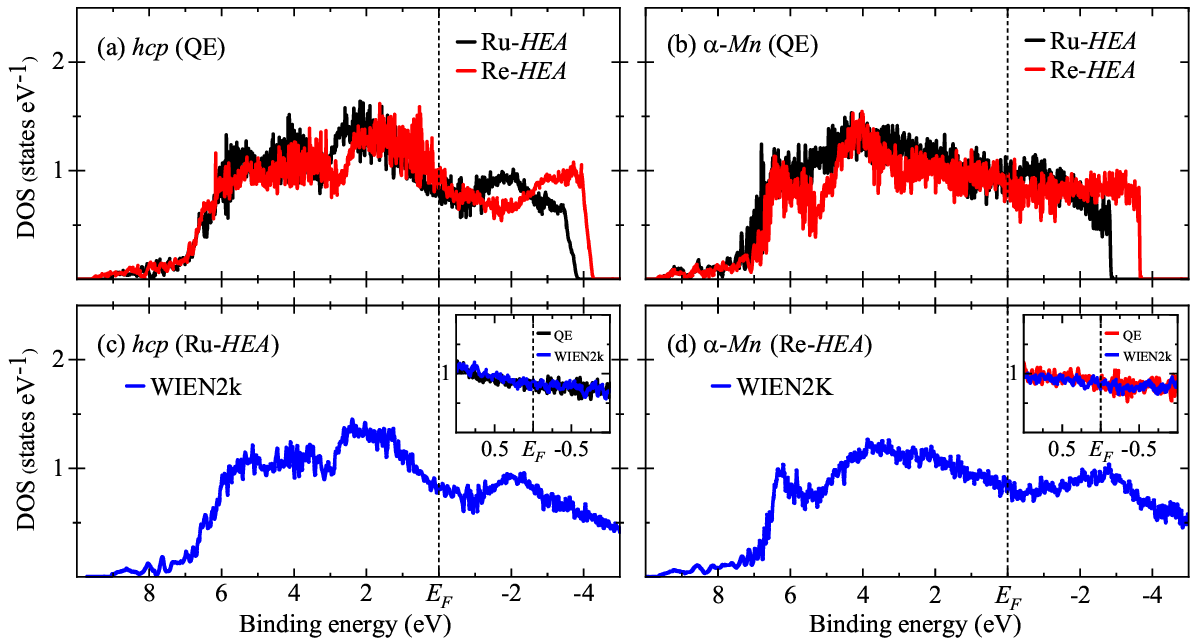}}
    \caption{Calculated DOS using QE for Ru-\textit{HEA} and Re-\textit{HEA} in (a) \textit{hcp} and (b) $\alpha$-\textit{Mn} structure. Comparison of DOS calculated using QE and Wien2k for (c) \textit{hcp}-Ru-\textit{HEA} and (d) $\alpha$-\textit{Mn}-Re-\textit{HEA} (Inset shows the DOS comparison near $E_F$). }\label{FigS3}
    \vspace{-2ex}
    \end{figure*}
    
\section{ Role of Structure and VEC on the Density of states}
To understand the observed differences in the DFT results presented in Fig. 2 of the main text for the two high entropy alloys, we show the results of DFT calculations for Re-\textit{HEA} in the \textit{hcp} structure and compare it with the results for Ru-\textit{HEA} (having experimental \textit{hcp} structure) as shown in Fig. \ref{FigS3}(a). The overall features of the valence band remain very similar in both the cases.  Clearly evident increased valence band width is primarily due to extended nature of 5$d$ orbitals while the smaller occupied density of states (DOS) (total states below $E_F$) is due to reduced valence electron count (VEC) (by 0.35 electrons) in case of Re-\textit{HEA} than those in case of Ru-\textit{HEA}. Similarly, as shown in Fig. \ref{FigS3}(b), we compare the results for Ru-\textit{HEA} in the $\alpha$-\textit{Mn} structure with Re-\textit{HEA} (having experimental $\alpha$-\textit{Mn} structure), exhibiting the role of VEC and extended nature of 5$d$ orbitals. These results confirm that the observed differences in the density of states (DOS) obtained for Ru-\textit{HEA} and Re-\textit{HEA} are primarily due to the difference in their crystal structure and VEC. To compare the results from pseudopotential-based calculation with an all-electron method, we computed the DOS using WIEN2k code \cite{10.1063/1.5143061} employing the local density approximation \cite{{PhysRevB.45.13244}} and a \textit{k}-mesh of 3$\times$5$\times$4 for Ru-\textit{HEA} and 5$\times$5$\times$5 for Re-\textit{HEA}. Figures \ref{FigS3}(c) and \ref{FigS3}(d) show the DOS calculated using WIEN2k for \textit{hcp}-structured Ru-\textit{HEA} and $\alpha$-\textit{Mn}-structured Re-\textit{HEA}, respectively. The WIEN2k results are very similar to those obtained with QE, reproducing the characteristic features and their energy positions. The difference appearing in the unoccupied region above $\sim$4 eV, where WIEN2k calculated DOS exhibits finite intensity which is absent in QE calculated DOS, arises from the unoccupied states those are not included in the pseudopotential generation for QE calculation. The insets show the comparison of DOS obtained using these two methods in the near $E_F$ region, exhibiting similarity with a decreasing trend at $E_F$.

 \section{ High resolution photoemission spectra}
 
\begin{figure}[t]
\centering
{\includegraphics[width=0.50\textwidth]{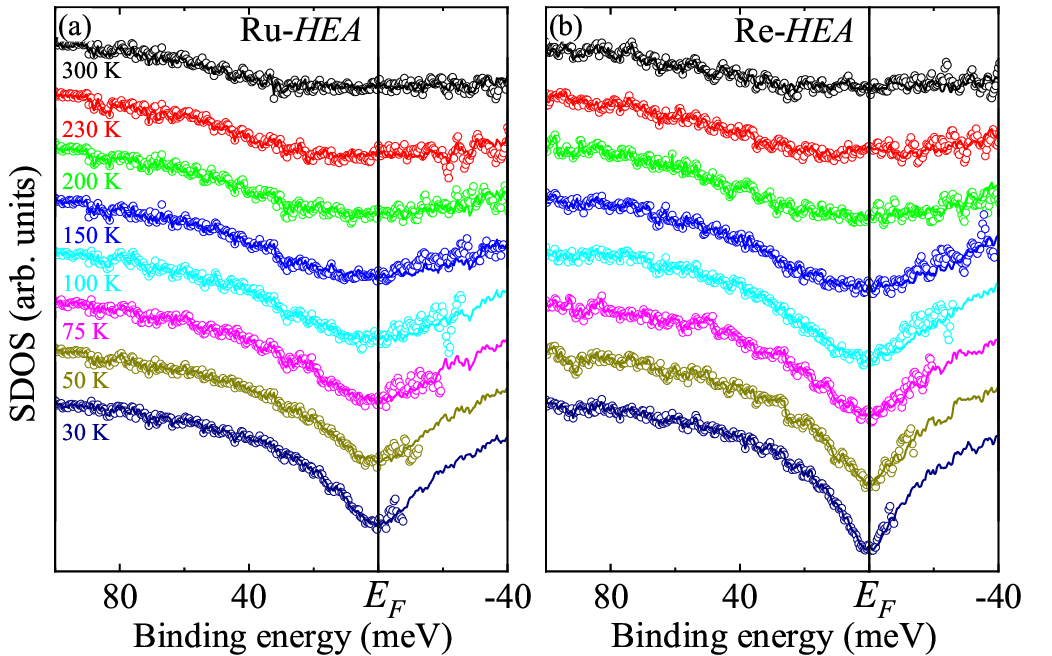}}
    \caption{Comparison of SDOS obtained using two methods for various temperatures for (a) Ru-\textit{HEA}  and (b) Re-\textit{HEA}. Open circles show the spectral DOS obtained by dividing the spectra with resolution broadened Fermi-Dirac function, and line represent the same obtained by symmetrization of the spectra.} \label{FigS4}
     \vspace{-2ex}
    \end{figure}

Figures \ref{FigS4}(a) and \ref{FigS4}(b) show the spectral DOS (SDOS) obtained using two different methodologies from high-resolution photoemission spectra collected using He {\scriptsize I} radiations at different temperatures for Ru-\textit{HEA} and Re-\textit{HEA} respectively. All the spectra are normalized at 200 meV BE. The photoemission intensity can be expressed as $I(E)$ = DOS($E$)$\ast$$F(E,T)$$\ast$$L^e(E)$$\ast$$L^h(E)$$\ast$$G(E)$, where $L^e$ ($L^h$) representing photoelectron (photohole) lifetime broadening. $F$ and $G$ represent Fermi-Dirac function and resolution broadening, respectively. Since, electron and hole lifetime broadenings ($L^e$ and $L^h$) are expected to be negligibly small in the close vicinity to the $E_F$, thus, $I(E)$/[$F(E,T)$$\ast$$G(E)$] provides a good representation of SDOS, where the photoemission intensity is divided by the resolution broadened Fermi-Dirac function at respective temperature $T$ \cite{PhysRevB.77.201102,PhysRevMaterials.7.064007,*Reddy_2019,PhysRevLett.95.016404,Bansal_2022}. Since, the Fermi-Dirac function (also resolution broadened) follows the relation $F(E)+F(-E) = 1$, the symmetrized photoemission intensity, $I(E)+I(-E)$, also provides a good description of SDOS provided that the SDOS near $E_F$ does not abruptly change \cite{PhysRevB.77.201102,PhysRevMaterials.7.064007,*Reddy_2019}. The solid lines represent the SDOS obtained by symmetrizing the photoemission intensity, while the overlaid scatter symbols correspond to the SDOS obtained by dividing the photoemission spectra by the resolution-broadened Fermi-Dirac function (plotted only up to 3$k_BT$ energy above $E_F$). All the spectra are stacked vertically for clearer visualization. Almost identical SDOS obtained by two different methods for both the systems provides confidence in the analysis. The SDOS in the vicinity of $E_F$ for both the high entropy alloys system, exhibit significant reduction in intensity with decreasing temperature due to strong disorder present in the systems.

\section{Electron-phonon coupling strength and $T_C$}

Calculating the electron-phonon coupling strength and $T_C$ for large-atom supercell is computationally prohibitive,
 whereas the VCA approach has been shown to reliably reproduce electron-phonon coupling and $T_C$ \cite{PhysRevLett.132.166002}, phonon band structure \cite{Körmann2017}, and mechanical and thermophysical properties \cite{HUANG2021109859}, particularly when the constituent elements occupy adjacent rows or columns of the periodic table. The VCA-calculated phonon band structure captures the low-frequency phonon behavior well, although it cannot account for disorder-induced broadening and tends to be less accurate for high-frequency dispersions \cite{Körmann2017}.

We computed the phonon dispersion curves and phonon linewidth representing lifetime broadening using DFPT within VCA for both the high entropy alloys in \textit{hcp} structure. The Eliashberg spectral function, electron-phonon ($e\text{-}ph$) coupling strength $\lambda$($\omega$) and $T_C$ were obtained from these calculations.

Eliashberg spectral function, $\alpha^2F(\omega)$ is defined as, 

\begin{equation}
\alpha^2F(\omega) = \frac{1}{2\pi N(E_F)} \sum_{\mathbf{q},\nu} \delta(\omega - \omega_{\mathbf{q}\nu}) \frac{\gamma_{\mathbf{q}\nu}}{\hbar \omega_{\mathbf{q}\nu}}
\end{equation}
where N($E_F$) is the DOS at $E_F$ and $\gamma_{\mathbf{q}\nu}$ is phonon linewidth for mode $\nu$ at wavevector q.

The cumulative $e\text{-}ph$ coupling strength, $\lambda$ is derived from Eliashberg spectral function using the equation 
\begin{equation}
\lambda = 2 \int_0^\infty \frac{\alpha^2F(\omega)}{\omega} \, d\omega
\end{equation}
and logarithmic average frequency $\omega_{\log}$ can be calculated by,

\begin{equation}
\omega_{\log} = \exp\left( \frac{2}{\lambda} \int_0^\infty \frac{d\omega}{\omega} \alpha^2F(\omega) \ln \omega \right)
\end{equation}

The $T_C$ was evaluated by the Allen-Dynes modified McMillan formula

\begin{equation}
T_c = \frac{\omega_{\log}}{1.2} \exp\left( \frac{-1.04(1 + \lambda)}{\lambda - \mu^* - 0.62 \lambda \mu^*} \right)
\end{equation}

The Coulomb pseudopotential $\mu^*$ (typically ranging from 0.10 to 0.16) was set to 0.13, as commonly used for high entropy alloys systems \cite{Motla_2023,Li_2025}. VCA calculations reveal suppression of the $T_C$ in \textit{hcp} phase of  Ru-\textit{HEA} and Re-\textit{HEA}, indicating a significant role of disorder in suppressing the $T_c$. 

In high entropy alloys phase stability is significantly influcend by VEC, which can also govern the formation of complex crystal structures \cite{doi:10.1021/ja00022a004,PhysRevB.109.024102,cryst7010009}. Within a given structure $T_C$ often correlates with VEC. Notably $T_C$ increases linearly with VEC in the $\alpha$-\textit{Mn} structure \cite{Li_2025} and a high Re content seems to be important for the stability of this phase.
To elucidate the impact of composition and VEC, we considered a hypothetical \textit{hcp} structured Re based high entropy alloy with the composition Re$_{0.15}$Os$_{0.55}$Mo$_{0.10}$W$_{0.10}$Zr$_{0.10}$. This configuration yields a VEC of 7.05, sufficiently higher than the $\alpha$-\textit{Mn} structured Re-\textit{HEA} and which may stabilize the \textit{hcp} phase. Our phonon calculations for this system give $\lambda = 0.74$ and $T_c$ = 5.96 K suggesting reduction of $T_C$ with increase in VEC for \textit{hcp} structure. These results demonstrate that compositional optimization, particularly through VEC tuning with controlled phase stabilization may collectively influence superconducting properties of these high entropy alloys.